\documentclass[twocolumn,9pt]{article}
\usepackage{amsmath,amssymb}
\usepackage{graphicx}
\usepackage{dcolumn}
\usepackage[algo2e,ruled,vlined]{algorithm2e}
\usepackage{hyperref}

\makeatletter
\newbox\abstract@box
\renewenvironment{abstract}
  {\global\setbox\abstract@box=\vbox\bgroup
     \hsize=\textwidth\linewidth=\textwidth
    \small
    \begin{center}%
    {\bfseries \abstractname\vspace{-.5em}\vspace{\z@}}%
    \end{center}%
    \quotation}
  {\endquotation\egroup}
\expandafter\def\expandafter\@maketitle\expandafter{\@maketitle
  \ifvoid\abstract@box\else\unvbox\abstract@box\if@twocolumn\vskip1.5em\fi\fi}
\makeatother

\title{An efficient implementation of Slater-Condon rules}

\begin{document}
\author{Anthony Scemama\thanks{
Laboratoire de Chimie et Physique Quantiques, CNRS-IRSAMC,
Universit\'e de Toulouse, France.},
Emmanuel Giner
}

    \begin{abstract}
Slater-Condon rules are at the heart of any quantum chemistry method as they
allow to simplify $3N$-dimensional integrals as sums of 3- or 6-dimensional
integrals. In this paper, we propose an efficient implementation of those rules
in order to identify very rapidly which integrals are involved in a matrix element
expressed in the determinant basis set. This implementation takes advantage of
the bit manipulation instructions on x86 architectures that were introduced in
2008 with the SSE4.2 instruction set. Finding which spin-orbitals are involved
in the calculation of a matrix element doesn't depend on the number of electrons
of the system.
\vspace{1cm}
    \end{abstract}

\maketitle



In this work we consider wave functions $\Psi$ expressed as linear combinations of
Slater determinants $D$ of orthonormal spin-orbitals $\phi({\bf r})$:
\begin{equation}
\Psi = \sum_i c_i D_i
\end{equation}
Using the Slater-Condon rules,\cite{slater,condon} the matrix elements of any one-body
(${\cal O}_1$) or two-body (${\cal O}_2$) operator expressed in the
determinant space have simple expressions involving one- and two-electron
integrals in the spin-orbital space.
The diagonal elements are given by:
\begin{eqnarray}
  \langle D | {\cal O}_1 | D \rangle & = & \sum_{i \in D} \langle \phi_i | {\cal O}_1 | \phi_i \rangle \\
  \langle D | {\cal O}_2 | D \rangle & = & \frac{1}{2} \sum_{(i,j) \in D}  
      \langle \phi_i \phi_j | {\cal O}_2 | \phi_i \phi_j \rangle - \nonumber \\
 & & 
      \langle \phi_i \phi_j | {\cal O}_2 | \phi_j \phi_i \rangle \nonumber 
\end{eqnarray}
For two determinants which differ only by the substitution of spin-orbital $i$ with
spin-orbital $j$:
\begin{eqnarray}
  \langle D | {\cal O}_1 | D_i^j \rangle & = & \langle \phi_i | {\cal O}_1 | \phi_j \rangle \\
  \langle D | {\cal O}_2 | D_i^j \rangle & = & \sum_{k \in D} 
      \langle \phi_i \phi_k | {\cal O}_2 | \phi_j \phi_k \rangle - 
      \langle \phi_i \phi_k | {\cal O}_2 | \phi_k \phi_j \rangle \nonumber
\end{eqnarray}
For two determinants which differ by two spin-orbitals:
\begin{eqnarray}
  \langle D | {\cal O}_1 | D_{ik}^{jl} \rangle & = & 0 \\
  \langle D | {\cal O}_2 | D_{ik}^{jl} \rangle & = & 
      \langle \phi_i \phi_k | {\cal O}_2 | \phi_j \phi_l \rangle -
      \langle \phi_i \phi_k | {\cal O}_2 | \phi_l \phi_j \rangle \nonumber 
\end{eqnarray}
All other matrix elements involving determinants with more than two
substitutions are zero. 

An efficient implementation of those rules requires:
\begin{enumerate}
 \item to find the number of spin-orbital substitutions between two determinants
 \item to find which spin-orbitals are involved in the substitution
 \item to compute the phase factor if a reordering of the spin-orbitals has occured
\end{enumerate}
This paper proposes an efficient implementation of those three points by using
some specific bit manipulation instructions at the CPU level.

\section{Algorithm}
\label{algorithm}

In this section, we use the convention that the least significant bit of binary
integers is the rightmost bit. As the position number of a bit in an integer is
the exponent for the corresponding bit weight in base 2, the bit positions are
numbered from the right to the left starting at position 0. To be consistent
with this convention, we also represent the arrays of 64-bit integers from
right to left, starting at position zero.
Following with the usual notations, the spin-orbitals start with index one.

\subsection{Binary representation of the determinants}

The molecular spin-orbitals in the determinants are ordered by spin: the
$\alpha$ spin-orbitals are placed before the $\beta$ spin-orbitals.  Each
determinant is represented as a pair of bit-strings: one bit-string
corresponding to the $\alpha$ spin-orbital occupations, and one bit-string for
the $\beta$ spin-orbital occupations.  When the $i$-th orbital is occupied by
an electron with spin $\sigma$ in the determinant, the bit at position $(i-1)$
of the $\sigma$ bit-string is set to one, otherwise it is set to zero.

The pair of bit-strings is encoded in a 2-dimensional array of 64-bit integers.
The first dimension contains $N_{\rm int}$ elements and starts at position zero.
$N_{\rm int}$ is the minimum number of 64-bit integers needed to encode the
bit-strings:
\begin{equation}
N_{\rm int} = \lfloor N_{\rm MOs}/64 \rfloor + 1
\end{equation}
where $N_{\rm MOs}$ is the total number of molecular spin-orbitals with spin
$\alpha$ or $\beta$ (we assume this number to be the same for both spins). The
second index of the array corresponds to the $\alpha$ or $\beta$ spin. Hence,
determinant $D_k$ is represented by an array of $N_{\rm int}$ 64-bit integers
$I_{i,\sigma}^k$, $i \in [0,N_{\rm int}-1], \sigma \in \{1,2\}$.

\subsection{Finding the number of substitutions}

\begin{algorithm2e}[t]
 \LinesNumbered

 \SetKwData{Degree}{d}
 \SetKwData{DegreeTwo}{two\_d}
 \SetKwFunction{Popcnt}{popcnt}
 \SetKwFunction{Xor}{~xor~}
 \SetKwFunction{And}{and}
 \SetKwFunction{Nsubst}{n\_excitations}
 
 \DontPrintSemicolon
 \SetKw{Function}{Function}
 \Function{\Nsubst{$I^1, I^2$}}\;
 \KwIn{$I^1$, $I^2$: lists of integers representing determinants $D_1$ and $D_2$.}
 \KwOut{\Degree : Degree of excitation.}
 \PrintSemicolon

 \DegreeTwo $\gets 0$\;
 \For {$\sigma \in \{\alpha, \beta \}$ } {
   \For {$i \gets 0$ \KwTo $N_{\rm int}-1$ } {
     $\DegreeTwo \gets \DegreeTwo +$ \Popcnt($I_{i,\sigma}^1 \Xor I_{i,\sigma}^2$)\;
   }
 }
 $\Degree \gets \DegreeTwo/2$\;
 \Return{$\Degree$}\;
\caption{Compute the degree of excitation between $D_1$ and $D_2$.}
\label{func:find_degree}
\end{algorithm2e}

We propose an algorithm to count the number of substitutions between two
determinants $D_1$ and $D_2$ (algorithm~\ref{func:find_degree}).  This number
is equivalent to the degree of excitation $d$ of the operator $\hat{T}_d$
which transforms $D_1$ into $D_2$ ($D_2 = \hat{T}_d D_1$).
The degree of excitation is equivalent to the number of holes created in $D_1$ or
to the number of particles created in $D_2$. In algorithm~\ref{func:find_degree},
the total number of substitutions is calculated as the sum of the number of
holes created in each 64-bit integer of $D_1$.

On line 4, $(I_{i,\sigma}^1 \Xor I_{i,\sigma}^2)$ returns a 64-bit integer with
bits set to one where the bits differ between $I_{i,\sigma}^1$ and
$I_{i,\sigma}^2$. Those correspond to the positions of the holes and the
particles. The {\tt popcnt} function returns the number of non-zero bits 
in the resulting integer.
At line 5, {\tt two\_d} contains the sum of the number of holes and particles,
so the excitation degree {\tt d} is half of {\tt two\_d}.

The Hamming weight is defined as the number of non-zero bits in a binary
integer.The fast calculation of Hamming weights is crucial in various domains
of computer science such as error-correcing codes\cite{hamming} or
cryptography\cite{crypto}. Therefore, the computation of Hamming weights
has appeared in the hardware of processors in 2008 via the 
the {\em popcnt} instruction introduced with the SSE 4.2 instruction set.
This instruction has  a 3-cycle latency and a 1-cycle throughput independently
of the number of bits set to one (here, independently of the number of
electrons), as opposed to Wegner's algorithm\cite{wegner} that repeatedly finds
and clears the last nonzero bit. The {\em popcnt} instruction may be generated
by Fortran compilers via the intrinsic {\tt popcnt} function.

\subsection{Identifying the substituted spin-orbitals}

\begin{algorithm2e}[t]
 \LinesNumbered

 \SetKwData{II}{i}
 \SetKwData{K}{k}
 \SetKwData{Hole}{H}
 \SetKwData{Position}{position}
 \SetKwArray{Holes}{Holes}
 \SetKwFunction{BSF}{trailing\_zeros}
 \SetKwFunction{ibclr}{bit\_clear}
 \SetKwFunction{Xor}{~xor~}
 \SetKwFunction{And}{~and~}
 \SetKwFunction{GetHoles}{get\_holes}
 \SetKw{Function}{Function}
 
 \Function{\GetHoles{$I^1$, $I^2$}}\;
 \KwIn{$I^1$, $I^2$: lists of integers representing determinants $D_1$ and $D_2$.}
 \KwOut{\Holes: List of positions of the holes.}

 \For {$\sigma \in \{\alpha, \beta \}$ } {
    \K $\gets 0$\;
    \For {$\II \gets 0$ \KwTo $N_{\rm int}-1$ } {
       $\Hole \gets \left( I_{\II,\sigma}^1 \Xor I_{\II,\sigma}^2 \right)$ \And $I_{\II,\sigma}^1$\;
       \While { $\Hole \ne 0$ } {
         \Position $\gets$ \BSF{\Hole}\;
         \Holes{$\K,\sigma$} $\gets 1+ 64 \times \II +\Position$\;
         $\Hole \gets \ibclr(\Hole,\Position)$\;
         $\K \gets \K +1$\;
     }
    }
  }
  \Return{\Holes}\;

 \caption{Obtain the list of orbital indices corresponding to holes in the excitation from $D_1$ to $D_2$}
 \label{func:find_holes}
\end{algorithm2e}

Algorithm~\ref{func:find_holes} creates the list of spin-orbital indices
containing the holes of the excitation from $D_1$ to $D_2$.
At line 4, $H$ is is set to a 64-bit integer with ones at the positions of the holes.
The loop starting at line 5 translates the positions of those bits to
spin-orbital indices as follows: when $H \ne 0$, the index of the rightmost bit of
$H$ set to one is equal to the number of trailing zeros of the integer.
This number can be obtained by the x86\_64 {\em bsf} (bit scan
forward) instruction with a latency of 3 cycles and a 1-cycle throughput, and
may be generated by the Fortran {\tt trailz} intrinsic function.
At line 7, the spin-orbital index is calculated.
At line 8, the rightmost bit set to one is cleared in $H$.

\begin{algorithm2e}[t]
 \LinesNumbered

 \SetKwData{KK}{k}
 \SetKwData{II}{i}
 \SetKwData{Particle}{P}
 \SetKwData{Position}{position}
 \SetKwArray{Particles}{Particles}
 \SetKwFunction{BSF}{trailing\_zeros}
 \SetKwFunction{ibclr}{bit\_clear}
 \SetKwFunction{Xor}{~xor~}
 \SetKwFunction{And}{~and~}
 \SetKwFunction{GetParticles}{get\_particles}
 
 \DontPrintSemicolon 
 \SetKw{Function}{Function}
 \Function{\GetParticles{$I^1$, $I^2$}}\;
 \KwIn{$I^1$, $I^2$: lists of integers representing determinants $D_1$ and $D_2$.}
 \KwOut{\Particles: List of positions of the particles.}
 \PrintSemicolon 

 \For {$\sigma \in \{\alpha, \beta \}$ } {
    \KK $\gets 0$\;
    \For {$\II \gets 0$ \KwTo $N_{\rm int}-1$ } {
       $\Particle \gets \left( I_{\II,\sigma}^1 \Xor I_{\II,\sigma}^2 \right)$ \And $I_{\II,\sigma}^2$\;
       \While { $\Particle \ne 0$ } {
         \Position $\gets$ \BSF{\Particle}\;
         \Particles{$\KK,\sigma$} $\gets$ 1 + $64\times \II$ + \Position \;
         $\Particle \gets \ibclr(\Particle,\Position)$\;
         $\KK \gets \KK +1$\;
     }
    }
  }
  \Return{\Particles}\;

  \caption{Obtain the list of orbital indices corresponding to particles in the excitation from $D_1$ to $D_2$}
  \label{func:find_particles}
\end{algorithm2e}

The list of particles is obtained in a similar way with
algorithm~\ref{func:find_particles}.

\subsection{Computing the phase}

\begin{algorithm2e}[t]
 \LinesNumbered

 \SetKwData{KK}{k}
 \SetKwData{II}{i}
 \SetKwData{Phase}{phase}
 \SetKwArray{Holes}{Holes}
 \SetKwArray{Particles}{Particles}
 \SetKwFunction{Xor}{~xor~}
 \SetKwFunction{And}{~and~}
 \SetKwFunction{Not}{not}
 \SetKwFunction{Degree}{n\_excitations}
 \SetKwFunction{GetPhase}{GetPhase}
 
 \DontPrintSemicolon 
 \SetKw{Function}{Function}
 \Function{\GetPhase{\Holes, \Particles}}\;
 \KwIn{\Holes and \Particles obtained with alorithms~\ref{func:find_holes} and~\ref{func:find_particles}.}
 \KwOut{\Phase $\in \{-1,1\}$.}
 \SetKw{KwRequire}{Requires:}
 \KwRequire{\Degree{$I^1, I^2$} $\in \{1,2\}$. \Holes and \Particles are sorted.} 
 \SetKwFunction{Popcnt}{popcnt}
 \SetKwFunction{And}{~and~}
 \SetKwFunction{Or}{~or~}

 \PrintSemicolon 
 \SetKwArray{Filter}{mask}
 \SetKwData{K}{k}
 \SetKwData{JJ}{j}
 \SetKwData{Low}{low}
 \SetKwData{High}{high}
 \SetKwData{MN}{m}
 \SetKwData{NN}{n}
 \SetKwData{II}{i}
 \SetKwData{LL}{l}
 \SetKwData{Nperm}{nperm}
 \SetKwData{One}{One}
 \SetKwArray{PhaseDble}{phase\_dble}
 
 $\Nperm \gets 0$\;
 \For {$\sigma \in \{\alpha, \beta\}$} {
   $n_\sigma \gets$ Number of excitations of spin $\sigma$\;
   \For {$\II \gets 0$ \KwTo $n_\sigma-1$} {
     $\High \gets \max ($\Particles{$\II,\sigma$}, \Holes{$\II,\sigma$}$)$\;
     $\Low \gets \min ($\Particles{$\II,\sigma$}, \Holes{$\II,\sigma$}$)$\;
     $\K \gets \lfloor \High/64 \rfloor $\;
     $\MN \gets \High \pmod{64}$\;
     $\JJ \gets \lfloor \Low /64 \rfloor $\;
     $\NN \gets \Low  \pmod{64}$\;
     \For {$\LL \gets j$ \KwTo $k-1$} {
       \Filter{\LL} $\gets$\Not{$0$}\;
     }
     \Filter{\KK} $\gets 2^{\MN}-1$\;
     \Filter{\JJ} $\gets$ \Filter{\JJ} \And $($\Not{$2^{\NN+1}$}$+1)$

     \For {$\LL \gets \JJ$ \KwTo \KK} {
       $\Nperm \gets \Nperm + $\Popcnt{$I_1^{\JJ,\sigma} \And $\Filter{\LL}}\;
     } 
   } 

   \If {$(n_\sigma = 2)$\And$($
   \Holes{$2,\sigma$} $<$ \Particles{$1,\sigma$} \Or 
   \Holes{$1,\sigma$} $>$ \Particles{$2,\sigma$} $)$}
   { $\Nperm \gets \Nperm + 1$\; }
   
 }
 \Return{$-1^{\Nperm}$}\;
 \caption{Compute the phase factor of $\langle D_1 | {\cal O} | D_2 \rangle$ }
 \label{func:phase}
\end{algorithm2e}

In our representation, the spin-orbitals are always ordered by increasing
index. Therefore, a reordering may occur during the spin-orbital substitution,
involving a possible change of the phase.

As no more than two substitutions between determinants $D_1$ and $D_2$ give a
non-zero matrix element, we only consider in algorithm~\ref{func:phase}
single and double substitutions. The phase is calculated as $-1^{N_{\rm
perm}}$, where $N_{\rm perm}$ is the number permutations necessary to bring the
spin-orbitals on which the holes are made to the positions of the particles.
This number is equal to the number of occupied spin-orbitals between these two
positions.

We create a bit mask to extract the occupied spin-orbitals placed between the
hole and the particle, and we count them using the {\tt popcnt} instruction. We
have to consider that the hole and the particle may or may not not belong to
the same 64-bit integer.

On lines 6 and 7, we identify the highest and lowest spin-orbitals involved in
the excitation to delimitate the range of the bit mask. Then, we find to which
64-bit integers they belong and what are their bit positions in the integers
(lines 8--11). The loop in lines 12--13 sets to one all the bits of the mask
contained in the integers between the integer containing the lowest orbital
(included) and the integer containing highest orbital (excluded). 
Line 14 sets all the $m$ rightmost bits of the integer to one and all the other
bits to zero.
At line 15, the $n+1$ rightmost bits of the integer containing to the lowest
orbital are set to zero.
At this point, the bit mask is defined on integers \Filter{j} to \Filter{k}
(if the substitution occurs on the same 64-bit integer, $j=k$). 
\Filter{j} has zeros on the leftmost bits and \Filter{k} has zeros on the
rightmost bits. We can now extract the spin-orbitals placed between the hole
and the particle by applying the mask and computing the Hamming weight of the
result (line 17).

For a double excitation, if the realization of the first excitation introduces
a new orbital between the hole and the particle of the second excitation
(crossing of the two excitations), an additional permutation is needed, as done
on lines 18--19. This {\em if} statement assumes that the Holes and Particles
arrays are sorted.

\section{Optimized Implementation}
\label{implementation}

In this section, we present our implementation in the Fortran language. In
Fortran, arrays start by default with index one as opposed to the convention
chosen in the algorithms.

\begin{figure}[t]
\centering
\includegraphics[width=\columnwidth]{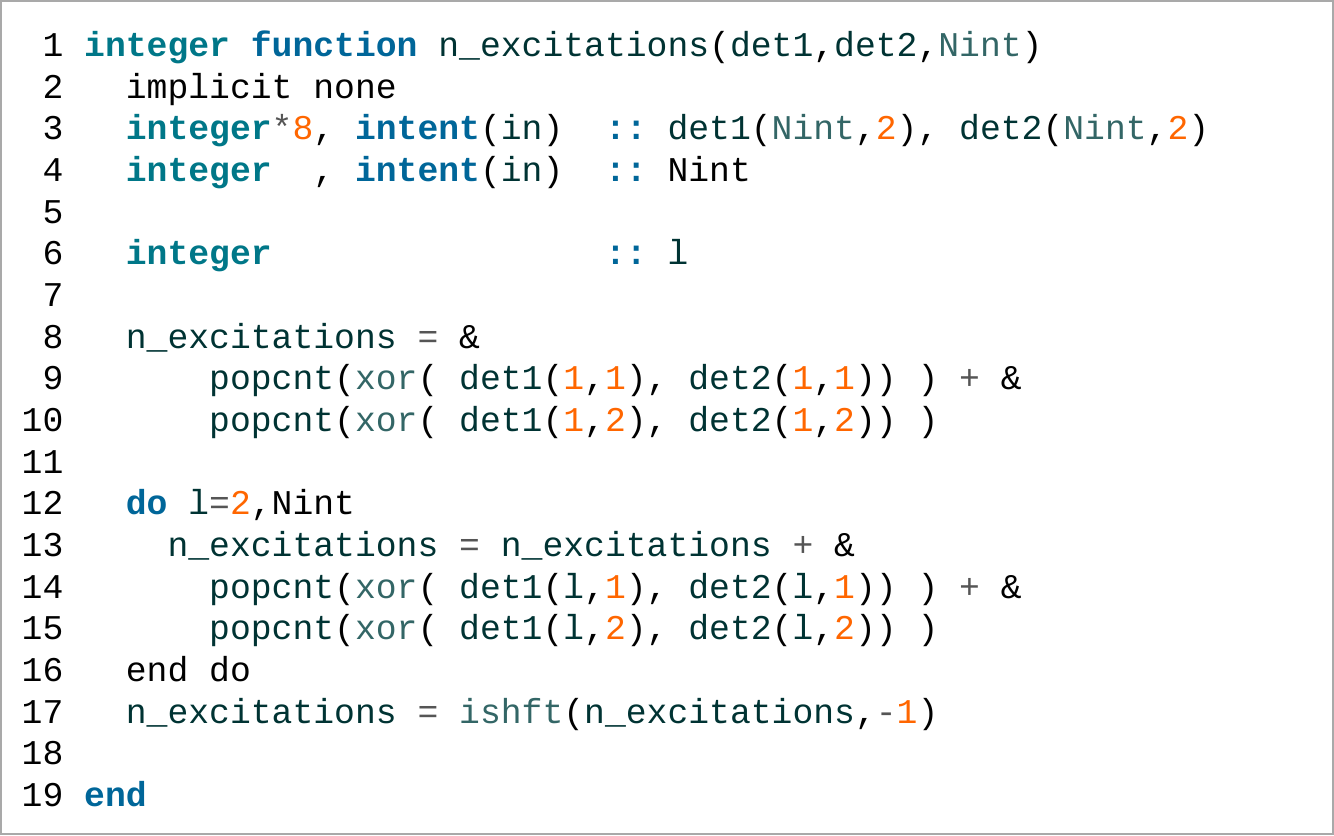}
\caption{Fortran implementation of algorithm~\ref{func:find_degree}.}
\label{fig:nsubst}
\end{figure}

A Fortran implementation of algorithm~\ref{func:find_degree} is given in
Figure~\ref{fig:nsubst}. This function was compiled with the Intel Fortran
compiler 14.0.0 with the options \verb=-xAVX -O2=.  A static analysis of the
executable was performed using the MAQAO tool\cite{maqao} : if all the data fit
into the L1 cache, an iteration of the loop takes in average 3.5~CPU cycles on
an Intel Sandy Bridge CPU core. A dynamic analysis revealed 10.5 cycles for
calling the function, performing the first statement and exiting the function,
and 4.1 CPU cycles per loop iteration. Therefore, to obtain the best
performance this function will need to be inlined by the compiler, eventually
using compiler directives or inter-procedural optimization flags.

\begin{figure}
\centering
\includegraphics[width=\columnwidth]{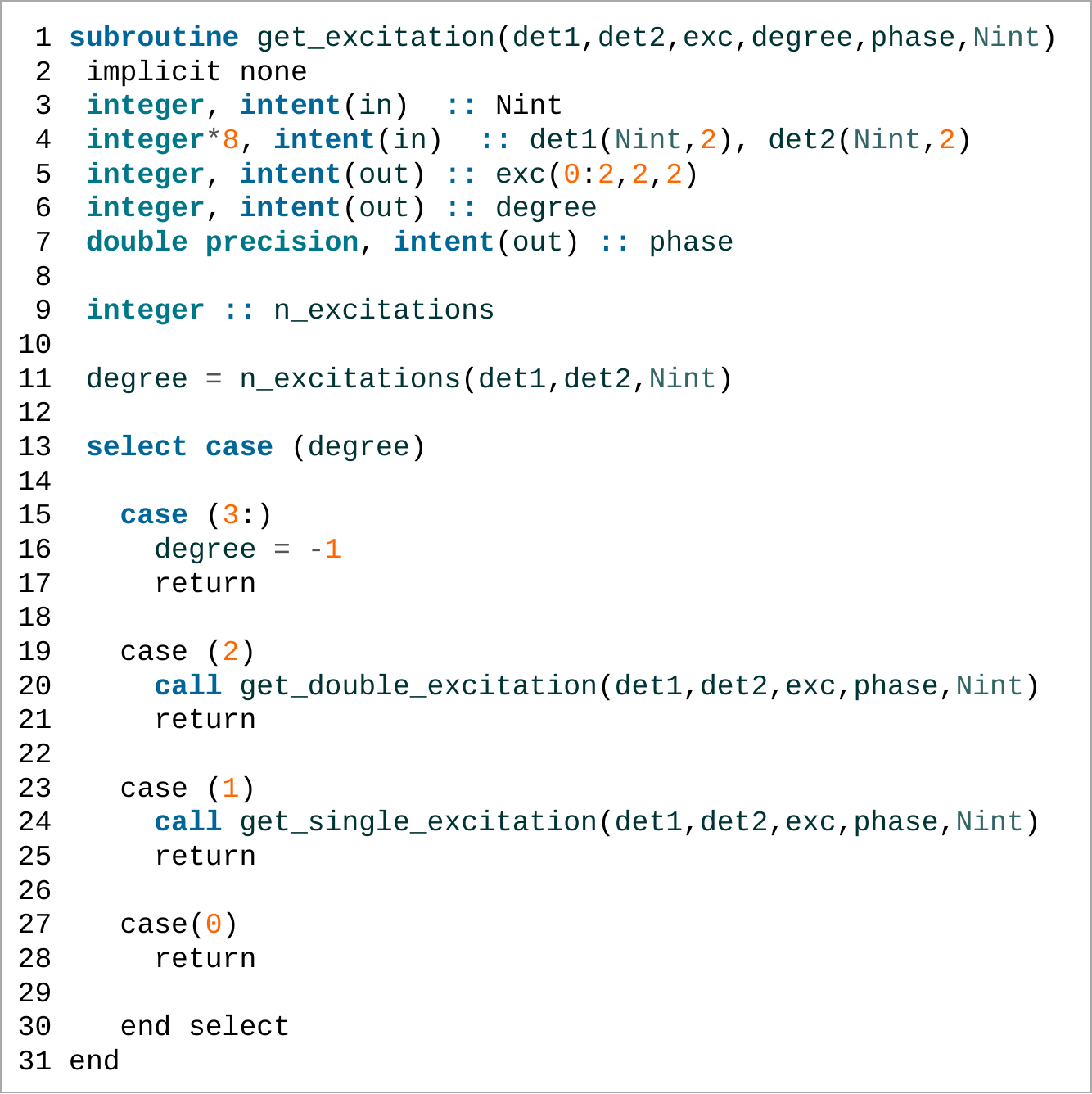}
\caption{Fortran subroutine for finding the holes and particles involved in a matrix element.}
\label{fig:exc0}
\end{figure}

For the identification of the substitutions, only four cases are possible
(figure~\ref{fig:exc0}) depending of the degree of excitation~:
no substitution, one substitution, two substitutions or more than two
substitutions.  If the determinants are the same, the subroutine exits with a
degree of excitation of zero.
For degrees of excitation higher than two, the subroutine returns a degrees of
excitation equal to -1.
For the two remaining cases, a particular subroutine is written for each case. 
The cases were ordered from the most probable to the least probable to optimize
the branch prediction.
In output, the indices of the spin-orbitals are given in the array {\tt exc} as
follows:
\begin{itemize}
 \item The last index of {\tt exc} is the spin (1 for $\alpha$ and 2 for $\beta$)
 \item The second index of {\tt exc} is 1 for holes and 2 for particles
 \item The element at index 0 of the first dimension of {\tt exc} gives the
total number of holes or particles of spin $\alpha$ or $\beta$
 \item The first index of {\tt exc} refers to the orbital index of the hole or particle
\end{itemize}

\begin{figure}
\centering
\includegraphics[width=\columnwidth]{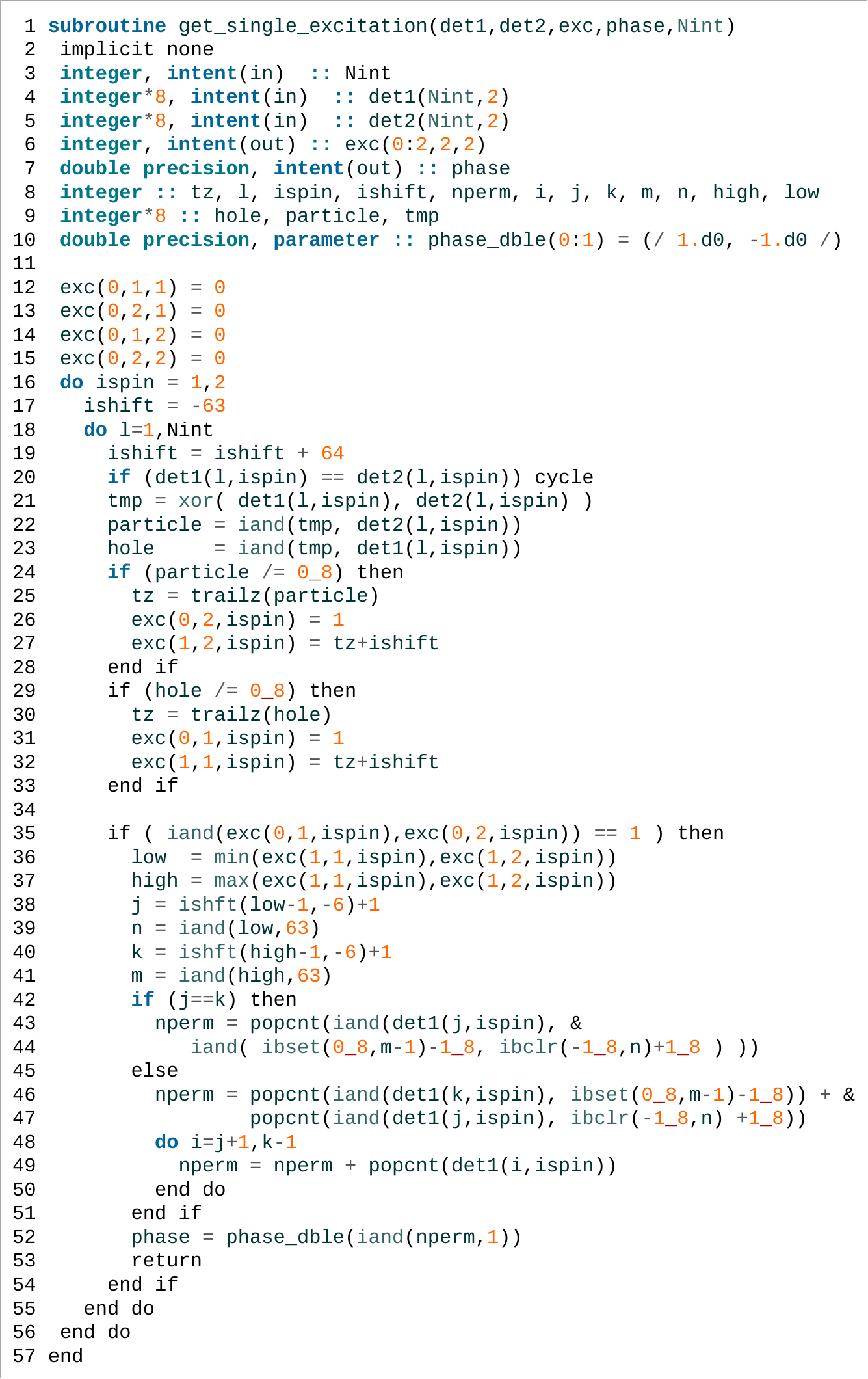}
\caption{Fortran subroutine for finding the holes and particles involved in a single excitation.}
\label{fig:exc1}
\end{figure}

The subroutine for single excitations is given in figure~\ref{fig:exc1}.
The particle and hole are searched simultaneously. 
The {\tt ishift} variable (line 19) allows to replace the integer
multiplication at line 7 of algorithm~\ref{func:find_holes} by an integer
addition, which is faster.
Line 38 contains a bit shift instruction where all the bits are
shifted 6 places to the right. This is equivalent to doing the integer division
of {\tt high} by 64. Line 39 computes {\tt high}$\pmod{64}$ using a bit mask.
The compiler may recognize that those last two optimizations are possible, but
it can not be aware that {\tt high} is always positive. Therefore, it will
generate additional instructions to handle negative integers. As we know
that {\tt high} is always positive, we can do better than the compiler.
The test at line 35 is
true when both the hole and the particle have been found. This allows to
compute the phase factor and exit the subroutine as soon at the single
excitation is obtained.
If the hole and particle belong to the same integer (line 42), the bit mask
is created and applied on the fly to only one 64-bit integer. Otherwise,
the bit mask is created and applied on the two extreme integers $j$ and $k$.
The integers between $j$ and $k$, if there are any, don't need to have a bit mask
applied since the mask would have all bits set to one.
Finally, line 52 calculates $-1^{N_{\rm perm}}$ using a memory access depending
on the parity of $N_{\rm perm}$: {\tt phase\_dble} is an array of two double
precision values (line 10).

For double excitations, the subroutine is similar to the previous one.
The {\tt nexc} variable counts how many holes and particles have been found,
in order to exit the loop (line 45) as soon as the double excitation is found.
The calculation of the phase is the same as the case of single excitations
(lines 48--67), but in the case of a double excitation of the same spin, 
orbital crossings can occur (lines 68--76).

\begin{figure}
\centering
\includegraphics[width=\columnwidth]{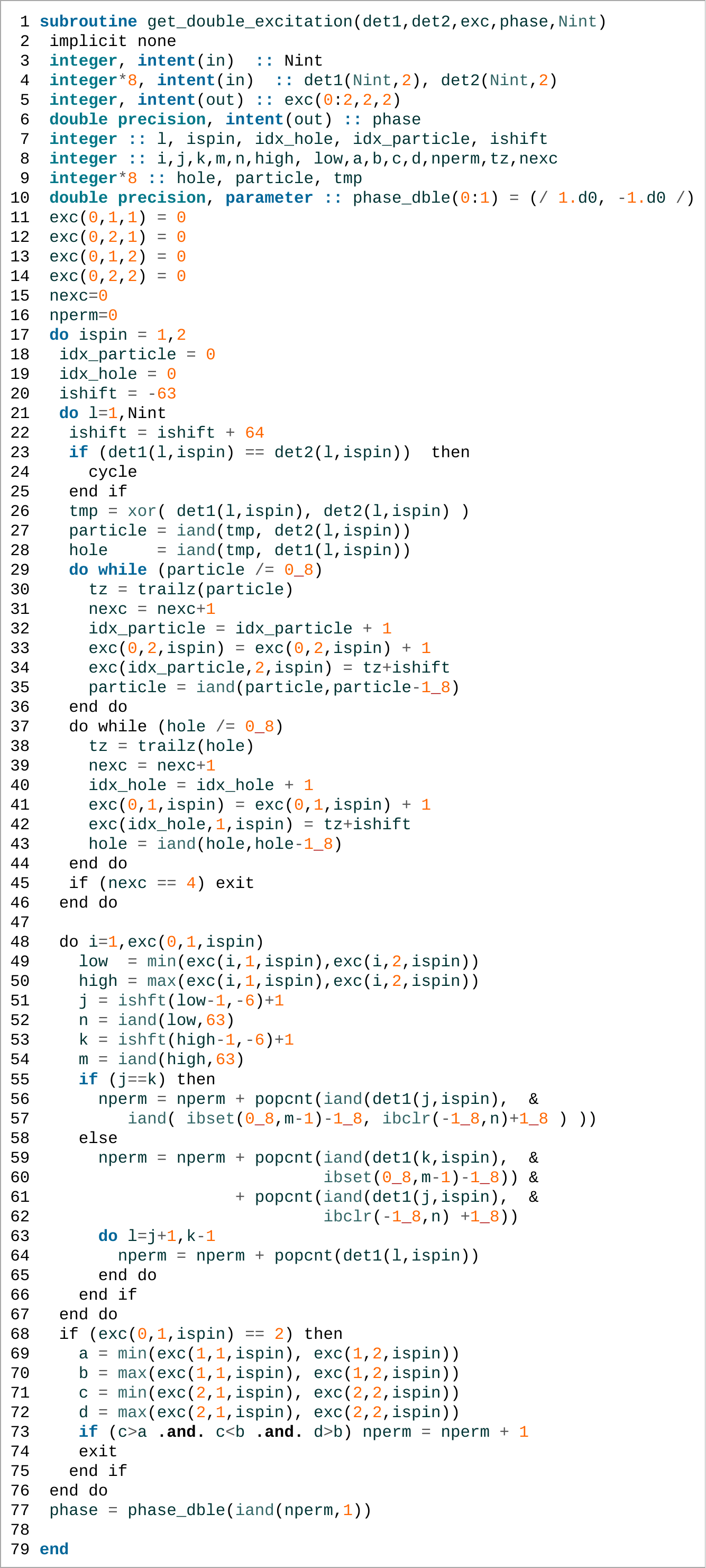}
\caption{Fortran subroutine for finding the holes and particles involved in a double excitation.}
\label{fig:exc2}
\end{figure}

\section{Benchmarks}
\label{benchmarks}

\begin{figure}[t]
\centering
\includegraphics[width=\columnwidth]{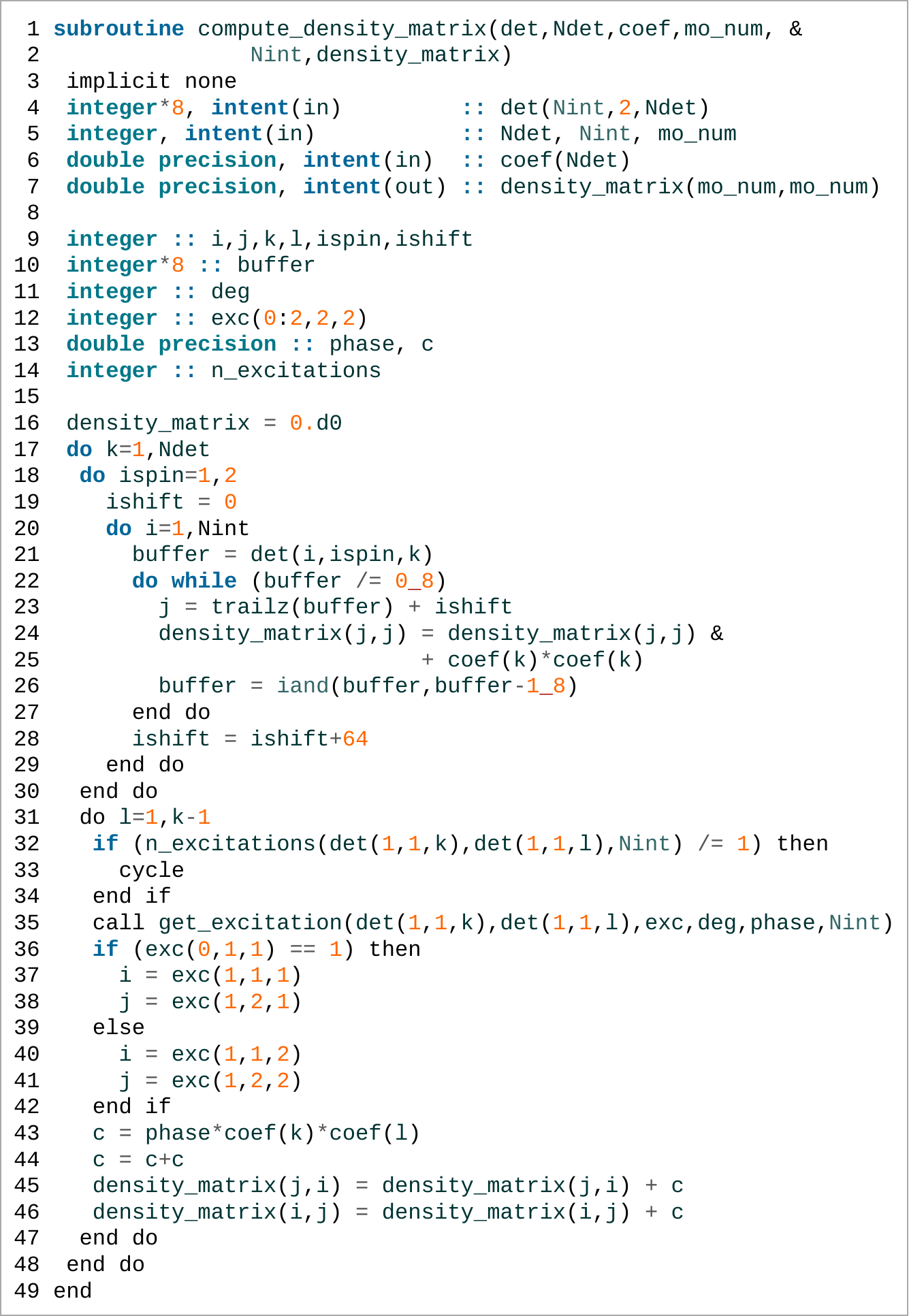}
\caption{Fortran subroutine for the calculation of the one-electron density matrix in the molecular orbital basis.}
\label{fig:dm}
\end{figure}

All the benchmarks were realized on a quad-core Intel Xeon CPU E3-1220 @ 3.10GHz
(8~MiB cache). The benchmark program is single-threaded, and was run using a
single CPU core at the maximum turbo frequency of 3.4~GHz. 
The numbers of CPU cycles were obtained by polling the hardware time stamp
counter {\tt rdtscp} using the following C function:
\begin{verbatim}
double rdtscp_(void) {
 unsigned long long a, d;
 __asm__ volatile ("rdtscp" : "=a" (a),
                              "=d" (d));
 return (double)((d<<32) + a);
}
\end{verbatim}

Two systems were benchmarked. Both systems are a set of 10~000 determinants obtained
with the CIPSI algorithm presented in ref~\cite{giner2013}.
The first system is a water molecule in the cc-pVTZ basis set\cite{ccpvtz}, in which
the determinants are made of 5 $\alpha$- and 5 $\beta$-electrons in 105
molecular orbitals ($N_{\rm int}=2$). The second system is a Copper atom
in the cc-pVDZ\cite{ccpvdz} basis set, in which the determinants are made of
15  $\alpha$- and 14 $\beta$-electrons in 49 molecular orbitals ($N_{\rm int}=1$).
The benchmark consists in comparing each determinant with all the derminants
($10^8$ determinant comparisons). These determinant comparisons are
central in determinant driven calculations, such as the calculation of the
Hamiltonian matrix in the determinant basis set. As an example of a practical
application, we benchmark the calculation of the one-electron density matrix on
the molecular orbital basis using the subroutine given in figure~\ref{fig:dm}.

The programs were compiled with the Intel Fortran Compiler
version 14.0.0.  The compiling options included inter-procedural optimization
to let the compiler inline functions via the {\tt -ipo} option, and the
instruction sets were specified using the {\tt -xAVX} or the {\tt -xSSE2}
options.
Let us recall that the AVX instruction set includes all the instructions
introduced with SSE4.2. Using AVX instructions, the {\tt popcnt} function is
executed through its hardware implementation, as opposed to SSE2 in which
{\tt popcnt} is executed through its software implementation.

In table~\ref{tab:results} we report measures of the CPU time and of the
average number of CPU cycles of all the subroutines presented in this paper
for the water molecule and the Copper atom.

The measures of {\tt n\_excitations} and {\tt get\_excitation} are made using
all the possible pairs of determinants.  Most of the time the degree of
excitation $d$ is greater than 2, so the average number of cycles has a large
weight on the $d>2$ case, and this explains why the average number of cycles is
much lower than in the $d=1$ and $d=2$ cases.

As the computation of the one-electron density matrix only requires to find the
spin-orbitals for the cases $d \in \{0,1\}$, which are a very small fraction of
the total, the average number of cycles is very close to this of {\tt
n\_excitations}. Note that the calculation of the density matrix only requires
$N(N+1)/2$ determinant comparisons, and this explains why the CPU time is
smaller than for the {\tt n\_excitations} benchmark.

All the source files needed to reproduce the benchmark presented in this section are
available at \url{https://github.com/scemama/slater_condon}.

\begin{table}[t]
\centering
\begin{tabular}{|l|D{.}{.}{1.2}D{.}{.}{1.2}|D{.}{.}{2.1}D{.}{.}{3.1}|}
\hline
                             &  \multicolumn{2}{c|}{Time (s)}           &     \multicolumn{2}{c|}{Cycles}     \\
                             &  \multicolumn{1}{c}{AVX} & \multicolumn{1}{c|}{\rm SSE2} & \multicolumn{1}{c}{\rm AVX} &  \multicolumn{1}{c|}{\rm SSE2}     \\
\hline                                                         
H$_2$O                       &                   &                      &                  &                   \\
{\tt n\_excitations}         &      0.33         &    2.33              &       10.2       &    72.6           \\
{\tt get\_excitation}        &      0.60         &    2.63              &       18.4       &    81.4           \\
{\tt get\_excitation}, $d=0$ &                   &                      &        6.2       &    58.7           \\
{\tt get\_excitation}, $d=1$ &                   &                      &       53.0       &   126.3           \\
{\tt get\_excitation}, $d=2$ &                   &                      &       88.9       &   195.5           \\
{\tt get\_excitation}, $d>2$ &                   &                      &        6.7       &    63.6           \\
Density matrix               &      0.19         &    1.23              &       11.7       &    75.7           \\
\hline
Cu                           &                   &                      &                  &                   \\
{\tt n\_excitations}         &      0.17         &    1.13              &        5.3       &    35.2           \\
{\tt get\_excitation}        &      0.28         &    1.27              &        8.7       &    39.1           \\
{\tt get\_excitation}, $d=0$ &                   &                      &        4.9       &    30.4           \\
{\tt get\_excitation}, $d=1$ &                   &                      &       47.0       &    88.1           \\
{\tt get\_excitation}, $d=2$ &                   &                      &       78.8       &   145.5           \\
{\tt get\_excitation}, $d>2$ &                   &                      &        5.5       &    29.3           \\
Density matrix               &      0.10         &    0.63              &        6.5       &    38.9           \\
\hline
\end{tabular}
\caption{CPU time (seconds) and average number of CPU cycles measured for the
{\tt n\_excitations} function, the {\tt get\_excitation} subroutine and the
calculation of the one-electron density matrix in the molecular orbital basis.  {\tt
get\_excitation} was also called using selected pairs of determinants such that
the degree of excitation ($d$) was zero, one, two or higher.}
\label{tab:results}
\end{table}

\section{Summary}
\label{conclusion}

We have presented an efficient implementation of Slater-Condon rules by taking
advantage of instructions recently introduced in x86\_64 processors. The use of
these instructions allow to gain a factor larger than 6 with respect to their
software implementation.  As a result, the computation of the degree of
excitation between two determinants can be performed in the order of 10 CPU
cycles in a set of 128 molecular orbitals, independently of the number of
electrons.  Obtaining the list of holes and particles involved in a single or
double excitation can be obtained in the order of 50--90 cycles, also
independently of the number of electrons. For comparison, the latency of a
double precision floating point division is typically 20--25 cycles, and a
random read in memory is 250--300 CPU cycles.  Therefore, the presented
implementation of Slater-Condon rules will significantly accelerate
determinant-driven calculations where the two-electron integrals have to be
fetched using random memory accesses. As a practical example, the one-electron density
matrix built from 10~000 determinants of a water molecule in the cc-pVTZ basis
set was computed in 0.2 seconds on a single CPU core.

\bibliography{slater_rules}
\bibliographystyle{unsrt}

\end{document}